\documentclass[10pt,a4paper]{article}

\usepackage{times,cite}
\usepackage{epsfig,wrapfig}
\usepackage{epstopdf} 
\usepackage{fancyhdr}
\usepackage[lmargin=2.5cm, rmargin=2.5cm,tmargin=3.50cm,bmargin=3.50cm]{geometry}
\usepackage{indentfirst}
\usepackage{titlesec}
\usepackage{subfigure}
\titleformat{\section}[hang]{\centering}{\thesection}{1ex}{\normalsize \textsc}
\titleformat{\subsection}[hang]{}{\thesubsection}{1ex}{\normalsize \textit}

\usepackage{epsfig}
\usepackage{graphicx}
\usepackage{subfigure}
\usepackage{color}
\usepackage{soul}
\usepackage{float}
\usepackage{epstopdf}
\usepackage{amsmath}
\usepackage{amssymb}
\usepackage{ifthen}
\usepackage{alltt}
\usepackage{fancyhdr}
\usepackage{verbatim}
\usepackage{moreverb}
\usepackage{rotating}
\usepackage{colortbl}
\usepackage{amssymb}
\usepackage{multirow}
\usepackage{bigstrut}
\usepackage{wrapfig}

\renewcommand{\thesection}{ \normalsize \textnormal{\Roman{section}.}}
\renewcommand{\thesubsection}{\normalsize \textnormal{\textsc{\textit{\Alph{subsection}.}}}}
\def\e{\begin{equation}}
\def\f{\end{equation}}
\def\_#1{{\bf #1}}

\def\.{\cdot}

\begin{document}
\title{\large \textbf{Metasurfaces for Perfect and Full Control of Refraction and Reflection}}
\def\affil#1{\begin{itemize} \item[] #1 \end{itemize}}
\author{\normalsize \bfseries V. Asadchy$^1$,   
    M. Albooyeh$^1$,
   S. Tcvetkova$^1$,
   Y. Ra'di$^{1,2}$,
     and
   \underline{S. A. Tretyakov}$^1$}
\date{}
\maketitle
\thispagestyle{fancy}
\vspace{-6ex}
\affil{\begin{center}\normalsize $^1$Department of Radio Science and Engineering, School of Electrical Engineering, Aalto University,\\
																      P.O. 13000, FI-00076, Aalto, Finland\\
$^2$Radiation Laboratory, Department of Electrical Engineering and Computer Science, University of Michigan, Ann Arbor, Michigan 48109, USA
\end{center}}

\begin{abstract}
\noindent \normalsize
\textbf{\textit{Abstract} \ \ -- \ \
In this talk  we present  and discuss a new general approach to the synthesis of metasurfaces for full control of  transmitted and reflected fields. The method is based on the use of an equivalent impedance matrix which connects the tangential field components at the two sides on the metasurface. Finding the impedance matrix components, we are able to synthesise metasurfaces which perfectly realize the desired response. We will explain possible alternative physical realizations and reveal the crucial role of bianisotropic coupling to achieve full control of transmission through perfectly matched metasurfaces. This abstract summarizes our results on metasurfaces for perfect refraction into an arbitrary direction. 
}\end{abstract}

\section{Introduction}

A metasurface is a composite material layer, designed
and optimized  to control and transform
electromagnetic waves. The layer thickness and the unit-cell size are small
as compared to the wavelength in the surrounding space. Recently, considerable efforts have been devoted to creation of metasurfaces for shaping reflected and transmitted waves, see e.g. review papers \cite{phil, Caloz_review,Elefth_review}. However, nearly always the realized properties of metasurfaces have not perfectly satisfied the design goals. In particular, in most attempts to synthesise metasurfaces for ``perfect refraction'' of plane waves travelling along a certain direction into a plane wave propagating along a different direction, it was either found that the surface must be active \cite{Tailoring} or some reflections are inevitable \cite{passive}. 

Here we summarize our recent results on synthesis of perfectly refractive  metasurfaces. Based on the general form of the impedance matrix of metasurfaces with the desired response, we identify a number of possible realizations of metasurfaces which exactly perform the desired operation. In the presentation, we will also discuss the  synthesis theory for full and perfect control of reflection.

\section{Requirements on the impedance matrix of perfectly refractive metasurfaces}

\begin{figure}[h!]
\centering
 \epsfig{file=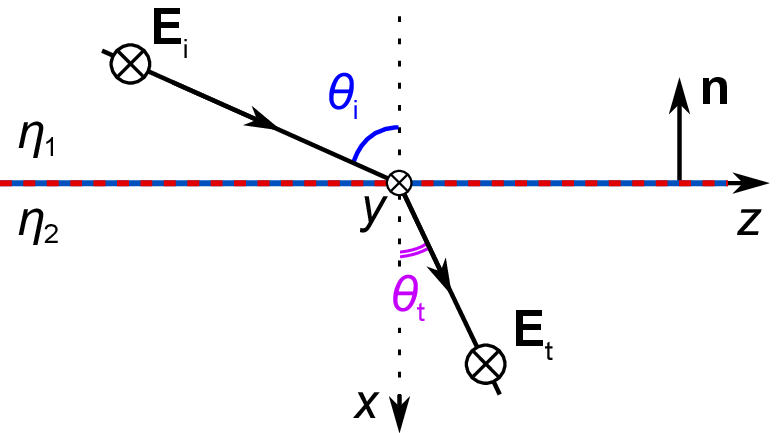, width=0.35\linewidth}  \hfill
 \epsfig{file=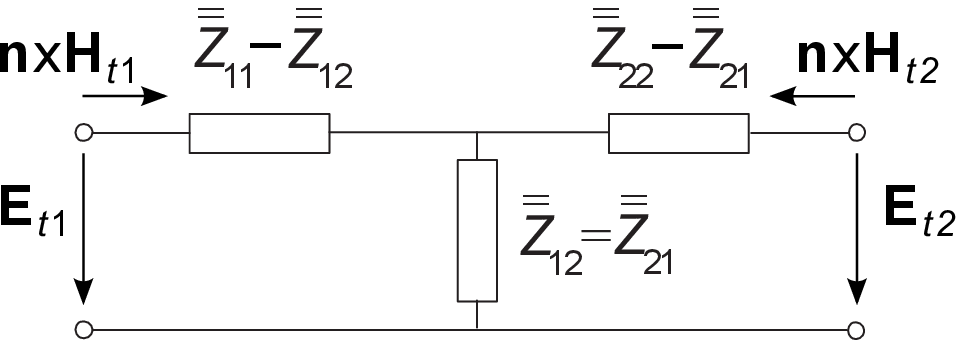, width=0.5\linewidth}  
  \caption{Illustration of the desired performance of an ideally refracting metasurface (left) and the equivalent $T$-circuit of a general reciprocal metasurface (right).}
  \label{geom}
\end{figure}

Let us set a goal to design a metasurface which perfectly (without reflection or absorption) refracts a plane wave in medium 1 (wave impedance $\eta_1$, wavenumber $k_1$, and the incidence angle $\theta_{\rm i}$) into a plane wave travelling in medium 2 (characterized by parameters $k_2$, $\eta_2$) in some other direction, specified by the refraction angle $\theta_{\rm t}$. The tangential field components $\_E_{t1}$ and $\_H_{t1}$  on the input side of the metasurface read (TE polarization as an example) 
\e \_E_{t1}=\_E_{\rm i}e^{-jk_1\sin\theta_{\rm i}z},\quad \_n\times \_H_{t1}=
\_E_{\rm i}{1\over \eta_1}\cos\theta_{\rm i}  e^{-jk_1\sin\theta_{\rm i}z}\label{plus} \f
where $z$ is the coordinate along the tangential component of the wavevector of the incident wave, and the unit vector $\_n$ is orthogonal to the metasurface plane, pointing towards the source (Fig.~\ref{geom}).
The required fields on the other side of the surface are 
\e \_E_{t2}=\_E_{\rm t}e^{-jk_2\sin\theta_{\rm t}z+j\phi_{\rm t}},\quad \_n\times \_H_{t2}=
\_E_{\rm t}{1\over \eta_2}\cos\theta_{\rm t}  e^{-jk_2\sin\theta_{\rm t}z+j\phi_{\rm t}} \label{minus}\f
($\phi_{\rm t}$ is the desired phase change in transmission). The synthesis method is based on the use of the impedance matrix, which relates the tangential fields on the two sides of the metasurface as 
\e  \_E_{t1}=Z_{11} \_n\times \_H_{t1} +Z_{12} (-\_n\times \_H_{t2}),\qquad  \_E_{t2}=Z_{21} \_n\times \_H_{t1} +Z_{22} (-\_n\times \_H_{t2})\label{21-22} \f

Substituting the required field values, we see that the metasurface will perform the desired operation exactly, if the impedance matrix components satisfy
\begin{equation}
e^{-jk_1\sin\theta_{\rm i}z}  = Z_{11}\, {1\over \eta_1}\cos\theta_{\rm i} \, e^{-jk_1\sin\theta_{\rm i}z} 
 -Z_{12}\, {1\over\sqrt{\eta_1 \eta_2}} \sqrt{\cos\theta_{\rm i} \cos\theta_{\rm t}} e^{-jk_2\sin\theta_{\rm t}z+j \phi_{\rm t}}\label{eq1}
\end{equation}
\begin{equation}
 e^{-jk_2\sin\theta_{\rm t}z+j \phi_t} =Z_{21}\, {1\over\sqrt{\eta_1 \eta_2}} \sqrt{\cos\theta_{\rm i} \cos\theta_{\rm t}}\, e^{-jk_1\sin\theta_{\rm i}z} -    Z_{22}\,  {\cos\theta_{\rm t}\over \eta_2}\, e^{-jk_2\sin\theta_{\rm t}z+j \phi_{\rm t}}\label{eq2a}
\end{equation}
Obviously, infinitely many alternative realizations are possible, because we need to satisfy only two complex equations for four complex parameters of the metasurface. 
We can impose additional conditions, for example, require some extra functionality or demand that the structure be made only from reciprocal or/and passive elements, etc. Some interesting possible approaches are discussed in the following.

\section{Possible realizations of perfectly refractive metasurfaces}

\noindent 1. {\bf Teleportation metasurface.}  
Inspecting (\ref{eq1}) and (\ref{eq2a}), we conclude that in general the metasurface should be non-uniform, as the parameters are functions of the coordinate $z$. However, there is an interesting  solution for which all the parameters are constants, and perfect refraction operation can be achieved using a uniform metasurface. Indeed, assuming that $Z_{12}=Z_{21}=0$, both equations are satisfied with 
\e Z_{11}={\eta_1\over \cos\theta_{\rm i}},\qquad Z_{22}=-{\eta_2\over \cos\theta_{\rm t}}\f   
In this scenario, the metasurface is formed by a matched absorbing layer (the input resistance $Z_{11})$, a perfect electric conductor  (PEC) sheet, and an active layer (an ``anti-absorber'' \cite{tele}) on the other side. The incident plane wave is totally absorbed in the matched absorber. The negative-resistance sheet (resistance $Z_{22}$) together with the wave impedance of medium~2 forms a self-oscillating system whose stable-generation regime corresponds to generation of a plane wave in the desired direction (the refraction angle $\theta_{\rm t}$): the sum of the wave impedance of plane waves propagating at the angle $\theta_{\rm t}$ and the input impedance of the active layer is zero, which is the necessary condition for stable generation.  This structure is similar to the ``teleportation metasurface'' introduced in \cite{tele} for teleportation of waves without changing their propagation direction.

\smallskip
\noindent 2. {\bf Transmitarray.}  
Alternatively, we can demand that the metasurface is matched for waves incident from medium~2 at the required refraction angle, which means that $Z_{22}={\eta_2\over \cos\theta_{\rm t}}$. Solving (\ref{eq2a}) with this value of $Z_{22}$, we find
\e Z_{21}={2\sqrt{\eta_1 \eta_2}\over  \sqrt{\cos\theta_{\rm i} \cos\theta_{\rm t}}}\, e^{-j(k_2\sin\theta_{\rm t}-k_1\sin\theta_{\rm i})z+j \phi_{\rm t}}\f
Furthermore, we can set $Z_{12}=0$ and $Z_{11}={\eta_1\over \cos\theta_{\rm i}}$, so that at the incidence angle $\theta_{\rm i}$ the metasurface is acting as a matched receiving antenna array. This nonreciprocal realization obviously corresponds to conventional transmitarrays (e.g. \cite{Zoya}). The incident plane wave is first received by a matched antenna array on one side of the surface and the wave is then launched into medium~2 with a transmitting phase array antenna. 

\smallskip
\noindent 3. {\bf Double current sheets.}
One can seek a realization in form of a metasurface which maintains both electric and magnetic surface currents, each subject to the corresponding impedance sheet conditions \cite{Caloz_review,Elefth_review,Tailoring,passive}
\e  \_J_{\rm e}=\_n\times (\_H_{t1}- \_H_{t2})=Y_{\rm e}\_E_t=Y_{\rm e}(\_E_{t1}+\_E_{t2})/ 2,\quad 
 \_J_{\rm m}=-\_n\times(\_E_{t1}-\_E_{t2})=Y_{\rm m}\_H_t=Y_{\rm m}(\_H_{t1}+ \_H_{t2})/2\label{Jm} \f  
Equations (\ref{plus}) tell that relations (\ref{Jm}) can hold only if the metasurface is symmetric and reciprocal, with  $Z_{11}=Z_{22}$  and  $Z_{12}=Z_{21}$. In this case, 
$ Y_{\rm e}=2/(Z_{11}+Z_{12})$,  $Y_{\rm m}=2(Z_{11}-Z_{12})$, and
solution of (\ref{eq1}) and (\ref{eq2a}) is unique. However, it gives complex-valued parameters \cite[suppl.~mat.]{Tailoring}, which means that the perfectly refractive double sheet must be lossy in some areas and active in some other areas of the surface. Purely reactive (lossless) parameters are possible only if the impedances seen from the opposite sides are equal, that is, $\eta_1/\cos\theta_{\rm i}=\eta_2/\cos\theta_{\rm t}$. This result lead to a conclusion  that lossless operation is possible only if there are some reflections \cite{passive}.

\smallskip
\noindent 4. {\bf Perfect metasurfaces formed by lossless elements.}  
However, perfect refraction by lossless reciprocal metasurfaces is possible if we allow bianisotropic coupling in the metasurface. It can be proved simply by solving (\ref{eq1}) and (\ref{eq2a})  in the assumption that the metasurface is lossless, that is, all $Z$-parameters are purely imaginary numbers. In this case the solution is again unique and reads
\e Z_{11}=-j{\eta_1\over \cos\theta_{\rm i}}\cot \alpha_{\rm t},\quad 
 Z_{22}=-j{\eta_2\over \cos\theta_{\rm t}}\cot \alpha_{\rm t}, \quad 
 Z_{12}=Z_{21}=-j{\sqrt{\eta_1 \eta_2}\over  \sqrt{\cos\theta_{\rm i} \cos\theta_{\rm t}} }{1\over \sin\alpha_{\rm t}}\label{X}\f
 where $\alpha_{\rm t}=(k_2\sin\theta_{\rm t}-k_1\sin\theta_{\rm i})z- \phi_{\rm t}$.
For the case of zero phase shift ($\phi_t=0$) formulas (\ref{X}) agree with  the recent result of \cite{Elefth_new}, obtained within the frame of the generalized scattering parameters  approach. 
Bianisotropic metasurfaces with the required properties defined by (\ref{X}) can be realized as arrays of low-loss particles with appropriate symmetry. For microwave applications, metallic canonical omega particles  or double arrays of patches (patches on the opposite sides of the substrate must be different to ensure proper magnetoelectric coupling) can be used. For optical applications, arrays of properly shaped dielectric particles were introduced as omega-type bianisotropic metasurfaces.

\section{Conclusion}

We have introduced a general approach to the synthesis of metasurfaces for arbitrary manipulations of plane waves. The main ideas of the method have been explained on an example of metasurfaces which perfectly refract plane waves incident at an angle $\theta_{\rm i}$ into plane waves propagating in an arbitrary direction defined by the refraction angle $\theta_{\rm t}$. The general synthesis approach shows a possibility for alternative physical realizations, and we have discussed several possible device realizations: self-oscillating teleportation metasurface, transmitarrays, double current sheets, and metasurfaces formed by only lossless components. The crucial role of omega-type bianisotropy in the design of lossless-component realizations has been revealed. Knowing $Z$-parameters (\ref{X}), it is easy to find the unit-cell polarizabilities or surface susceptibilities and optimize the unit-cell dimensions.
The method can be used also for synthesis of perfectly reflecting metasurface and for other transformations of waves (focusing, etc.) 

\end{document}